\documentclass[twocolumn,prb,showpacs]{revtex4}
\usepackage{graphicx}
\usepackage{dcolumn}
\usepackage{bm}

\begin {document}

\title {Influence of Anomalous Dispersion on Optical
Characteristics of Quantum Wells}
\author { L.I.Korovin, I.G.Lang}
\affiliation{A. F. Ioffe Physical-Technical Institute, Russian
Academy of Sciences, 194021 St. Petersburg, Russia}
\author{S.T.Pavlov\dag\ddag}
\address{\dag Facultad de Fisica de la UAZ, Apartado Postal C-580, 98060 Zacatecas, Zac., Mexico \\
\ddag P. N. Lebedev Physical Institute, Russian Academy of
Sciences, 119991 Moscow, Russia}

\begin {abstract}
Frequency dependencies of optical characteristics (reflection,
transmission and absorption of light) of a quantum well are
investigated in a vicinity of interband resonant transitions in a
case of two closely located excited energy levels. A wide quantum
well in a quantizing magnetic field directed normally to the
quantum-well plane, and monochromatic stimulating light are
considered. Distinctions between refraction coefficients of
barriers and quantum well, and a spatial dispersion of the light
wave are taken into account. It is shown that at large radiative
lifetimes of excited states in comparison with nonradiative
lifetimes, the frequency dependence of the light reflection
coefficient in the vicinity of resonant interband transitions is
defined basically by a curve, similar to the curve of the
anomalous dispersion of the refraction coefficient. The
contribution of this curve weakens at alignment of radiative and
nonradiative times, it is practically imperceptible at opposite
ratio of lifetimes . It is shown also that the frequency
dependencies similar to the anomalous dispersion do not arise in
transmission and absorption coefficients.
\end{abstract}


\maketitle

 Optical methods are widely used at research of electronic properties of low-dimensional semiconductor systems
  during last decades \cite{1,2,3,4}. It is
connected mainly that after interactions with such system, the
light wave contains an information about electronic processes, in
particular, about an electronic spectrum, lifetimes
 of excited states and scattering mechanisms.
 Interesting results turn out, when energy levels  of the electronic system are
discrete that takes place in quantum dots and quantum wells.
Discreteness of energy levels in a quantum well is provided by
excitonic states (if incident light is  perpendicular to the
quantum-well plane), or by quantizing magnetic field directed
perpendicularly to the quantum-well plane. In high-quality quantum
wells, a radiative broadening of absorption lines at low
temperatures and weak doping can be comparable with the
contribution of nonradiative relaxation mechanisms or to exceed
them. In this case, it is impossible to be limited by linear
approximation on interaction of electrons with electromagnetic
field, and it is necessary to consider all the orders of this
interaction
\cite{5,6,7,8,9,10,11,12,13,14,15,16,17,18,19,20,21,22,23,24}.
 Reflection, absorption and transmission of an
electromagnetic wave, which interacts with discrete energy levels
of electronic system in a quantum well in the region of
frequencies corresponding to interband transitions, were
considered in \cite{13,14,15,16,17,18,19}. In these works, a
stimulating light could be monochromatic \cite{19} or pulse
irradiation \cite{13,14,15}. One \cite{16}, two \cite{17, 19} and
large number of excited energy levels \cite{18} were considered.
Results of these works are applicable for narrow quantum wells,
when the inequality
\begin {equation}
\label {eq1} \kappa d \ll 1
\end {equation}
is carried out, where $d $ is the quantum well width, $ \kappa $
is the module of the light wave vector. Actually above mentioned
works are true in a zero approximation on the parameter $ \kappa d
$.

On the other hand, for wide quantum wells the parameter $ \kappa d
$ can be $ \approx 1 $. For example, for the GaAs hetero-laser
radiation (wave length $0.8 \mu $) and a quantum-well width $d
\cong 500 \AA $, the parameter $ \kappa d \cong 1.5 $. In this
case, it is necessary to take into account the spatial dispersion
of the electromagnetic wave, since its amplitude strongly varies
inside of the quantum well. Besides, for wide quantum wells, an
inequality $d \gg a _ {\, 0 \,} $ ($a _ {\, 0 \,} $ is the lattice
constant) is very strong, that allows
 to use Maxwell equations for continuous matters at determination of
 the electromagnetic field. Such approach allows to take into account
distinction in refraction coefficients of barriers and quantum
wells.

In \cite{20,21}, the theory considering spatial dispersion of an
electromagnetic wave at its passage through a quantum well is
developed. One excited energy level was considered (i.e., one
interband transition) and alongside with the spatial dispersion,
different refraction coefficients of barriers and of quantum well
were introduced for monochromatic \cite{20} and for pulse
\cite{21} irradiation. Work \cite{22} is devoted to the account of
a spatial dispersion of an electromagnetic wave in the case of two
closely located interband resonant transitions. That corresponds
to the magnetopolaron state in a quantum well \cite{23}.

At calculation of optical characteristics of a quantum well  in
the resonance, the contribution of resonant transitions is
allocated from dielectric permeability. It is considered
separately. These resonant transitions lead to occurrence of a
high-frequency current, in which alongside with the contribution,
frequency dependence of which corresponds to absorption, there is
the contribution, frequency dependence of which is similar to a
curve of the anomalous dispersion of the refraction coefficient.
Below, the role of these two contributions in formation of
frequency dependencies of the light reflection, transmission and
absorption by a quantum well is investigated as an example of two
closely located resonant transitions. It is generalization of two
previous works of authors: Unlike of \cite{20}, two excited energy
levels are taken into account, and results of \cite{22} are
generalized on a case, when refraction coefficients of barriers
and of quantum well are different.

\section{Basic relationships}

The system consisting of a semiconductor quantum well, located in
an interval $0 \leq z \leq d $, and of two semi-infinite barriers
is considered. A constant quantizing magnetic field is directed
perpendicularly to the quantum-well plane (along the $z $ axis).
The external plane electromagnetic wave extends along the $z $
axis from negative $z $. It is considered that barriers are
transparent for the electromagnetic wave, but in the quantum well
the wave is absorbed, causing resonant interband transitions. Zero
temperatures are assumed,
 a valence band is filled completely and a conduction band
is empty. In linear approximation on a wave amplitude, the excited
states are excitons. The frequencies of light close to the energy
gap are considered, when a small part of electrons of the valence
band takes part in light absorption. These electrons are located
near the band gap and the effective mass method is applicable for
them. For deep quantum wells in this case, it is possible to
neglect by electron tunneling into barriers. Besides, the energy
levels close to the quantum-well bottom can be considered in
approximation of infinitely high barriers. This restriction is not
principal and the theory can be extended on quantum wells of final
depths.

Since the size of heterogeneity (what is the quantum well) is
comparable with the light wavelength, optical characteristics of
such systems must be determined from the corresponding Maxwell
equation, where a current density is obtained on the basis of
microscopic consideration.

In the theory, interband matrix elements $ {\bf p} _ {\, c \, v} $
of quasi-momentum operator are essential. These matrix elements
correspond to the direct interband transition, i.e., to creation
of an electron-hole pair with coinciding coordinates of an
electron and hole. As well as in previous works \cite{18, 19, 24},
 the model is used, in which the vector $ {\bf p} _ {\, c \, v}
$ for two types of excitations $I $ and $II $ looks like:
\begin {eqnarray}
\label {eq2} {\bf p} _ { c  v \, I} &=& \frac {p _ { c  v}} {\sqrt
{2}} ({\bf e} _ { x }- i {\bf e} _ { y }),\nonumber\\ {\bf p} _
{\, c \, v \, II} &=& \frac {p _ {\, c \, v}} {\sqrt {2}} ({\bf e}
_ { x } + i {\bf e} _ { y }),
\end {eqnarray}
where $ {\bf e} _ {\, x \, (y)} $ is the unite vector along the $x
 (y)$ axis, $ p _ {\, c \, v} $ is some real constant. This model
corresponds to heavy holes in semiconductors with the zinc blend
structure, if the $z $ axis is directed along the 4th order
symmetry axis \cite{25, 26}. If the vectors of circular
polarization of stimulating light
\begin {equation}
\label {eq3} \bf e _ {\, \ell} = \frac {1} {\sqrt {2}}({\bf e} _
{\, x \,} \pm i \, {\bf e} _ {\, y \,})
\end {equation}
are introduced, the property of preservation of a polarization
vector is carried out. Thus, wave functions of the electron-hole
pair and energy levels do not depend on indices $I $ and $II $.

As is known, the electron-hole pair in a magnetic field is free,
if two conditions are satisfied. First, the Lorentz's force must
be large in comparison with Coulomb and exchange forces of
electron - hole interactions in pairs. Then, the wave function of
an electron-hole pair can be represented in the form of a product
of two functions: $ \Phi (z) $ and the function depending on $
{\bf r} _ {\, \perp} $ in the $xy $ plane of the quantum well.
Estimations show \cite{27, 28} that in GaAs for a magnetic field
corresponding to formation of magnetopolarons,
 this condition is carried out. The second
condition is some excess of a size-quantization energy  over the
Coulomb and exchange energy of interaction in the electron-hole
pair. Then, it is possible to consider the electron-hole pair as
free and to use the approximation of infinitely high barriers,
which is used below. The wave function describing dependence on $z
$ is as follows:
\begin {equation}
\label {eq4} \Phi _ {\lambda } (z) = {2\over d} \sin \Big({\pi
 m _c z\over d}\Big) \sin \Big({\pi
 m _v z\over d}\Big);~~~~ 0
\leq z \leq d
\end {equation}
and $ \Phi _ {\lambda} (z) =0 $ in barriers. The index $ \lambda =
(m _ { c }, m _ { v}) $ depends on quantum numbers of
size-quantization of electrons $m _ c $ and holes $m _ v $.

Let us note that in GaAs at $ \kappa d \geq 1 $, the second
condition of existence of free electron-hole pairs is not carried
out and the
 function $ \Phi (z) $  will be another there. However, the
approximation (4), essentially simplifying calculations, does not
affect qualitatively (as it is shown in Section 4) on frequency
dependence of optical characteristics of a quantum well.

In the present work, monochromatic light of frequency  $ \omega _
{\, \ell \,} $ and normal incidence on a quantum-well plane is
considered. According to these assumptions, an electric field of
stimulating light wave looks like:
\begin {equation}
\label {eq5} {\bf E} _ { 0 } = {\bf e} _ {\ell} E _ { 0 } e^ {-i
(\omega _ {\ell} t - \kappa _ {1 } z)} + ~c. c., \quad \kappa _ {
1 } = \nu _ {1 } \omega _ {\ell }/c,
\end {equation}
where $E _ {\, 0 \,} \, \, $ is the complex amplitude, $ \nu _ { 1
}  $ is the real material refraction coefficient in barriers. It
is supposed also that there are two closely located excited energy
levels in a quantum well, and the others excited energy levels in
the quantum well are located far enough. Such situation is
possible, for example, for a magnetopolaron state \cite{19, 23}.

In case of a direct interband transition, the average induced
charge density in a quantum well $ \rho (z, t) =0 $, that allows
to introduce the gauge $ \varphi (z, t) =0 $, where $ \varphi (z,
t) $ is the scalar potential. Then, the amplitude of electric
field in barriers is determined by the equation
\begin {equation}
\label {eq6} \frac {d ^{ 2 }  E} {dz ^{ 2 }} + \kappa _ { 1 }^ { 2
}  E=0, \kappa _ { 1 } = \frac {\nu _ { 1 } \omega _ { \ell }}
{c};~~~~ z \leq 0,  z \geq d,
\end {equation}
and in a quantum well $ {(0 \leq z \leq d)} $, it is determined by
the equation
\begin {equation}
\label {eq7} \frac {d^ { 2 }  E} {dz^ { 2 }} + \kappa^ { 2 }  E =
- \frac {4 \pi i \omega _ { \ell }} {c ^{ 2 }} {\bar J} (z);~~~~
\kappa = \frac {\nu \omega _ { \ell }} {c}.
\end {equation}
$c $ is the light velocity in vacuum; $ \nu _ { 1 },  \nu $ are
the refraction coefficient in barriers and in the quantum well,
respectively. $ {\bar J} (z) $ is the the Fourier-component of the
current density (averaged on the system ground state), which is
induced in the quantum well by a monochromatic plane wave. In case
of two excited energy levels, $ {\bar J} (z) $ for a quantum well
with infinitely high barriers looks like (The general formula is
represented, for example, in \cite{20, 24}.):
\begin {eqnarray}
\label {eq8} {\bar J} (z) = \frac {i \nu c} {4 \pi} \sum _ {j=1}
^2 \frac {\gamma _ { r  j } \Phi _ { j } (z)} {{\tilde \omega} _ {
j }} \int_0 ^{ d } dz ^{ ' } \Phi _ { j
} (z { ' }) E (z ^{ ' }),\nonumber\\
 0 \leq z \leq d
\end {eqnarray}
and ${\bar J} (z)  =  0$  in barriers. Here, $ \gamma _ { r  j } $
  is the radiative broadening of the excited states of an energy
doublet in the case of narrow quantum wells.

If the doublet is formed by a magnetopolaron $A $ to which the
Landau quantum number of a hole $n=1 $ (according to the
magnetopolaron classification \cite{23}), then,
\begin {equation}
\label {eq9} \gamma _ { r  j } = \gamma _ { r }  Q _ { 0  j},
\end {equation}
where
\begin {equation}
\label {eq10} \gamma _ { r } = \frac {2e^ { 2 }} {\hbar c \nu}
\frac {p _ { c  v }^ 2} {m _ { 0 } \hbar \omega _ { g }} \frac
{|e|H} {m _ { 0 } c}
\end {equation}
($ \hbar \omega _ { g } $ is the energy gap, $m _ { 0 } $ is the
free electron mass, $H $ is the magnetic field, $e $ is the
electron charge). The factor
\begin {eqnarray}
\label {eq11} Q _ { 0  j } &=& \frac {1} {2} \bigg (1 \pm {\lambda
\over \sqrt {\lambda^ { 2 } + (\Delta E _ {pol})^2}} \bigg),\nonumber\\
 \lambda &=& \hbar (\Omega _ { c } -\omega _ { LO})
\end {eqnarray}
determines a change of radiative lifetime at deviation of the
magnetic fields from the resonant value defined by equality $
\Omega _ { c } = \omega _ { LO} $. $ \Delta E _ { pol} $ is the
polaron splitting \cite{24}, $ \Omega _ { c } , \omega _ { LO} $
are the cyclotron frequency and longitudinal optical phonon
frequency, respectively. In the resonance, $ \lambda=0, Q _ { 0 j}
=1/2 $ and $ \gamma _ { r  1 } = \gamma _ { r  2 } $.

Resonant denominators in (8) are:
\begin {equation}
\label {eq12} {\tilde \omega} _ { j } = \omega _ { \ell } - \omega
_ { j } +i \gamma _ { j }/2,
\end {equation}
where $ \omega _ { j } $ are the frequencies of resonant
transitions of the doublet energy levels, $ \gamma _ {j} $ are the
radiative broadenings of these levels. In (8), resonant
denominators are taken into account only. Subscripts $j=1 $ and
$j=2 $ of the function $ \Phi _ { j } (z) $ correspond to pairs of
quantum numbers of the size-quantization, between which there is a
transition. $m _ { c } {( 1)}, m _ { v } {( 1)} $ correspond to
subscript $j=1 $; $m _ { c } {( 2)}, m _ { v } {( 2)} $ correspond
to subscript $j=2 $. The Landau quantum number $n $ is kept at the
direct transition.  The total field $E $ enters on the right-hand
side of the equation (8), that is connected with refusal from the
perturbation theory on the coupling constant $e ^{ 2 } / \hbar c
$.

\section{Electric field of an electromagnetic wave}

The further calculation is spent in the assumption that
\begin {eqnarray}
\label {eq13} m _ { c } {( 1)} &=&m _ { c } {( 2)} =m _ { c
},\nonumber\\ m _ { v } {( 1)} &=&m _ { v } {( 2)} =m _ { v },
\end {eqnarray}
that, in particular, corresponds to the magnetopolaron $A $. In
this case, the equality
\begin {equation}
\label {eq14} \Phi _ { 1 } (z) = \Phi _ { 2 } (z) = \Phi _ { m _ {
c }  m _ { v }} (z) \equiv \Phi (z)
\end {equation}
takes place and the formula (8) becomes
\begin {equation}
\label {eq15} {\bar J} (z) = \frac {i \nu c} {4 \pi} \bigg (\frac
{\gamma _ { r  1 }} {{\tilde \omega} _ { 1 }} + \frac {\gamma _ {
r  2 }} {{\tilde \omega} _ { 2 }} \bigg) \Phi (z) \int _ {0}^ { d}
d  z { ' } \Phi (z { ' }) E (z { ' }).
\end {equation}
The solution of the equation (6), determining the amplitude of the
field $E (z) $ in barriers, is
\begin {equation}
\label {eq16} E { \ell } (z) =E _ { 0 } e^ {i \kappa _ { 1 }  z} +
C _ { R } e ^{-i \kappa _ { 1 }  z};~~ z \leq 0,
\end {equation}
\begin {equation}
\label {eq17} E { r } (z) =C _ { T } e ^{i \kappa _ { 1 } z};~~ z
\geq d,
\end {equation}
$C _ { R } $ determines the amplitude of the reflected wave, $C _
{ T } $ determines the amplitude of the wave which passed through
the quantum well. It is convenient to represent the
integro-differential equation (7) for the field amplitude in a
quantum well in the form of the Fredholm integral equation of the
second type \cite{20}
\begin {eqnarray}
\label {eq18} E (z) &=&C _ { 1 } e ^{i \kappa  z} + C _ { 2 } e^
{-i \kappa  z}\nonumber\\&-& \frac {i} {2} \bigg (\frac {\gamma _
{ r 1 }} {{\tilde \omega} _ { 1 }} + \frac {\gamma _ { r  2 }}
{{\tilde \omega} _ { 2 }} \bigg) F (z) \int_0^ { d} d z { ' } \Phi
(z { ' }) E (z { ' }),
\end {eqnarray}
where
\begin {eqnarray}
\label {eq19} F (z) =F _ {m _ { c } m _ { v }} =e^ {i \kappa z}
\int_0 ^{ z} d  z ^{ ' } e ^{-i \kappa  z^ { ' }} \Phi (z ^{ '
})\nonumber\\ + e^ {-i \kappa z} \int_z ^{ d} d  z ^{ ' } e^ {i
\kappa  z^ { ' }} \Phi (z^ { ' }).
\end {eqnarray}
For arbitrary $m _ { c } $ and $m _ { v } $, $F (z) $ is equal
\begin {eqnarray}
\label {eq20} F (z) &=&i {\cal B} \bigg \{d  {\tilde \Phi} (z)-e^
{i \kappa  z}- (-1)^ {m _ { c } + m _ { v }} e^ {i \kappa
(d-z)}\nonumber\\ &+& \frac {d} {2} \bigg [ \frac {m _ { c }^ { 2}
+m _ { v }^ { 2}} {m _ { c } m _ { v }}- \frac {(\kappa  d)^ { 2}}
{\pi ^{ 2} m _ { c } m _ { v }} \bigg] \Phi (z) \bigg \}.
\end {eqnarray}
$ \Phi (z) $ is defined in (4),
\begin {eqnarray}
\label {eq21}
 {\tilde \Phi (z)} &=& {2\over d} \cos \Big({\pi  m _ { c } z\over d }\Big)\cos \Big({\pi  m _ { v } z\over d }\Big),
 \nonumber\\
 {\cal B} &=& 4 \pi^ { 2} m _ { c } m _ {
v } \kappa  d\nonumber\\&\times&
 [\pi ^{ 2} (m _ { c } + m
_ { v })^ { 2} - (\kappa  d)^ { 2}]^{-1}\nonumber\\&\times&
[(\kappa
 d)^ { 2}- \pi^ { 2} (m _ { c }-m _ { v })^ {
2}]^{-1}.
\end {eqnarray}
It follows from (19) and (20) that
\begin {eqnarray}
\label {eq22} F (0) &=&i {\cal B}  \big [1- (-1)^ {m _ { c } + m _
{ v }} e^ {i  \kappa  d} \big],\nonumber\\ F (d)& =& (-1)^ {m _ {
c } + m _ { v }}  F (0).
\end {eqnarray}

 Multiplying the equation (18) on $ \Phi
(z) $ and integrating on $z $ from $0 $ up to $d $, we obtain
\begin {equation}
\label {eq23} \int_0^ { d} d  z \Phi (z)  E (z) = \frac {h {\tilde
\omega} _ { 1 }  {\tilde \omega} _ { 2 }} {{\tilde \omega} _ { 1 }
 {\tilde \omega} _ { 2 } + (i   \varepsilon /2)
[\gamma _ { r  1}  {\tilde \omega} _ { 2 } + \gamma _ { r  2 }
{\tilde \omega} _ { 1 }]},
\end {equation}
where designations are introduced:
\begin {equation}
\label {eq24} \varepsilon = \varepsilon { ' } + i \varepsilon { "
} = \int_0 ^{ d} d  z  \Phi (z)  F (z),
\end {equation}
\begin {eqnarray}
\label {eq25} h &=& \int_0 ^{ d} d  z  \Phi (z) \big (C _ { 1 } e^
{i \kappa  z} + C _ { 2 }  e^ {-i \kappa  z} \big)\nonumber\\
&=&  F (0)  \big [C _ { 1 } + (-1)^ {m _ { c } + m _ { v }} e^ {-i
 \kappa  d}  C_2 \big].
\end {eqnarray}
As a result, complex amplitude of the electric field in a quantum
well becomes
\begin {eqnarray}
\label {eq26} E (z) &=&C _ { 1 }  e^{i \kappa  z} +C _ { 2 }  e^
{-i \kappa  z}\nonumber\\ &-& \frac {(i/2) h  F (z) (\gamma _ { r
1} {\tilde \omega} _ { 2 } + \gamma _ { r  2 }  {\tilde \omega} _
{ 1 })} {{\tilde \omega} _ { 1 }  {\tilde \omega} _ { 2 } + i
(\varepsilon/2) (\gamma _ { r  1}  {\tilde \omega} _ { 2 } +
\gamma _ { r  2 }  {\tilde \omega} _ { 1 })}.
\end {eqnarray}

Parameter $ \varepsilon $, as well as in the case of one excited
energy level, determines renormalization of the radiative
broadening $ \varepsilon { '} $ and shift $ \varepsilon { "} $ for
each of two excited levels. It follows from (24) and (20) that
\begin {eqnarray}
\label {eq27}
 Re  \varepsilon &=& \varepsilon { '} =
2 {\cal B} { 2 }  [1- (-1)^ {m _ { c } + m _ { v }}
\cos \kappa  d],\nonumber\\
 Im  \varepsilon &=& \varepsilon { "}\nonumber\\ &=&2 {\cal B}  \Bigg \{
\frac {(1 + \delta _ {m _ { c } m _ { v }}) (m _ { c } + m _ { v
})^ { 2 } + (m _ { c }-m _ { v })^ { 2 }} {8m _ { c } m _ { v
}}\nonumber\\&-&
 (-1)^{m _ { c } + m _ { v }} {\cal B} \sin
\kappa  d- \frac {(2 + \delta _ {m { c } m { v }}) (\kappa  d)^ {
2 }} {8m _ { c } m _ { v }} \Bigg \}
\end {eqnarray}
(At $ \kappa d \to 0 ~~~  \varepsilon^ { '} \to 1,   \varepsilon^
{ "} \to 0~~~   (m _ { c } = m _ { v }) $ and $ \varepsilon ^{ '}
\to 0,
 \varepsilon ^{ "} \to 0 ~~~  (m _ { c } \neq m _ { v
}) $). Thus, the real radiative broadening of energy levels of the
doublet is determined by values
\begin {equation}
\label {eq28} \varepsilon^{ '}  \gamma _ { r  i } = {\tilde
\gamma} _ { r  i },~~~~ i=1,2.
\end {equation}
That fact that $ \varepsilon ^{ '} $ and $ \varepsilon ^{ "} $ are
identical for both energy levels of the doublet is connected with
the assumption of equality of the size-quantization quantum
numbers $m _ { c  (v) }^ {( 1 )} = m _ { c  (v) }^ {( 2 )} $.

Arbitrary constants $C _ { 1 } $ and $C _ { 2 } $ enter, according
to (25), in the function $h $. Constants $C _ { 1 }, C _ { 2 }, C
_ { R } $ and $C _ { T } $ were determined from continuity
conditions of the magnetic field and the tangential projection of
the electric field on borders $z=0 $ and $z=d $. Normal
projections of an electric field are equal to zero. Arbitrary
constants are equal:
\begin {eqnarray}
\label {eq29} C _ { 1 }& = &\frac {2E _ { 0 }} {\Delta} e ^{-i
\kappa  d} [1 + \zeta + (1 - \zeta) N],\nonumber\\ C _ { 2 } &=& -
\frac {2E _ { 0 }} {\Delta} (1 - \zeta) \big [e ^{i \kappa  d} +
(-1)^ {m _ { c } + m _ { v }}  N \big],
\end {eqnarray}
\begin {eqnarray}
\label {eq30} C _ { R}& =& {E _ { 0 } \rho \over \Delta},~~~ C _ {
T } =
 {4E _ { 0 }\over \Delta } \nonumber\\
&\times& \zeta  e^ {-i \kappa _ { 1 }  d} \bigg [1 + (-1)^ {m _ {
c } + m _ { v }}  e^ {-i \kappa d}
 N \bigg] ,
\end {eqnarray}
\begin {eqnarray}
\label {eq31} \Delta &=& (\zeta +1)^ { 2}  e^ {-i \kappa  d} -
(\zeta-1)^ { 2}  e^ {i \kappa  d}- 2 (\zeta-1)\nonumber\\&\times&
\big [(\zeta +1) e^ {-i \kappa  d} + (-1)^ {m _ { c } + m _ { v }}
(\zeta-1) \big] N.
\end {eqnarray}
\begin {eqnarray}
\label {eq32} \rho&=&2i (\zeta ^{ 2}-1) \sin {\kappa d}\nonumber\\
&+&2 \big [(\zeta^ { 2} +1)  e^ {-i \kappa  d} + (-1)^ {m _ { c }
+ m _ { v }} (\zeta^ { 2}-1) \big]  N.
\end {eqnarray}
In (29) - (32), the designation
\begin {equation}
\label {eq33} \zeta = \kappa / \kappa _ { 1 } = \nu / \nu _ { 1 }
\end {equation}
is introduced and the function $N $ looks like:
\begin {eqnarray}
\label {eq34} N &=&-i (-1)^ {m _ { c } + m _ { v }} e ^{i \kappa
d}\nonumber\\&\times& \frac {({\tilde \gamma} _ {r1} {\tilde
\omega} _2 + {\tilde \gamma} _ {r2} {\tilde \omega} _1)}
{2[{\tilde \omega} _1 {\tilde \omega} _2 +i (\varepsilon/2)
(\gamma _ {r1} {\tilde \omega} _2 + \gamma _ {r2} {\tilde \omega}
_1)]}.
\end {eqnarray}
At obtaining of this formula, the equality
\begin {equation}
\label {eq35} F^2 (0)  = (-1)^ {m_c+m_v} e^ {i \kappa d}
\varepsilon '.
\end {equation}
is used.

 Uncertain coefficients (29) - (30) on their
form coincide with obtained in \cite{20} for one excited energy
level. Distinction consists in the function $N $, which in case of
$ \gamma _ { r  1 } $ (or $ \gamma _ { r  2 }) \to 0 $ passes in $
{\cal N} $ (the formula (41) in \cite{20}).

The curve $N (\omega _ { \ell }) $ is a function with two extrema,
etch of which corresponds to the resonant transition. It may be
represented more evidently, namely,
\begin {eqnarray}
\label {eq36} N & =& -i (-1)^ {m _ { c } + m _ { v }} e^ {i \kappa
d}\nonumber\\
&\times&\Bigg [ \frac {{\tilde \gamma} _ {r1}/2} {\Omega_1+i
\Gamma_1/2} +  \frac {{\tilde \gamma} _ {r2}/2} {\Omega_2+i
\Gamma_2/2} \Bigg].
\end {eqnarray}

Here, the renormalized resonant frequencies $ \Omega_1 $ and $
\Omega_2 $ and total broadening of each energy level $ \Gamma_1 $
and $ \Gamma_2 $ are equal:
\begin {eqnarray}
\label {eq37} \Omega_j &=& \omega_ \ell - \omega_j - \varepsilon^
{\prime \prime} \gamma _ {rj}/2- \beta_j ^{\prime
\prime},\nonumber\\ \Gamma_j &=& \gamma_j + {\tilde \gamma} _ {rj}
+2 \beta ' _j.
\end {eqnarray}
Values $ \beta _ {1 (2)}^ \prime $ and $ \beta _ {1 (2)}^ {\prime
\prime} $ make complex parameters
\begin {eqnarray}
\label {eq38} \beta _ { 1 } = \beta _ { 1 }^ { '} + i \beta _ { 1
}^ { "} = \frac {\varepsilon  \gamma _ { r  2 }} {2}  \frac
{{\tilde
\omega} _ { 1 }} {{\tilde \omega} _ { 2 }},\nonumber\\
\beta _ { 2 } = \beta _ { 2 }^ { '} + i \beta _ { 2 }^ { "} =
\frac {\varepsilon  \gamma _ { r  1 }} {2}  \frac {{\tilde \omega}
_ { 2 }} {{\tilde \omega} _ { 1 }}.
\end {eqnarray}
They determine mutual influence of energy levels. In the function
(36), two peaks are presented in the explicit form, but they are
not of the Lorentz-type, since parameters $ \beta_1 $ and $
\beta_2 $ ( bringing the contribution in broadening, and in shift
of peaks) depend on frequency. It is naturally that representation
of the function $N $ in the form of the sum Lorentzian's (as it is
made in \cite{19, 22}) and non-Lorentzian's curves leads to
identical results. Let us represent also an obvious form of
parameters $ \beta _ { j } ^{ '} $ and $ \beta _ { j }^ { "} $:

$$ \beta _ { 1 }^ { '} = \frac {({\tilde \gamma} _ { r  2 }/2) \sigma _ { 1 }-
\varepsilon^ { "} (\gamma _ { r  2 }/2) \sigma _ { 2 }} {(\omega _
{ \ell } - \omega _ { 2 })^ { 2} + \gamma _ { 2 }^ { 2}/4},$$$$
\beta _ { 2 }^ { '} = \frac {({\tilde \gamma} _ { r  1 }/2) \sigma
_ { 1 } + \varepsilon^ { "} (\gamma _ { r  1 }/2) \sigma _ { 2 }}
{(\omega _ { \ell } - \omega _ { 1 })^ { 2} + \gamma _ { 1 }^ {
2}/4}, $$
$$ \beta _ { 1 }^ { "} = \frac {({\tilde \gamma} _ { r  2 }/2) \sigma _ { 2 } +
\varepsilon^ { "} (\gamma _ { r  2 }/2) \sigma _ { 1 }} {(\omega _
{ \ell } - \omega _ { 2 }) ^{ 2} + \gamma _ { 2 }^ { 2}/4},$$$$
\beta _ { 2 }^ { "} = \frac {-({\tilde \gamma} _ { r  1 }/2)
\sigma _ { 2 } + \varepsilon ^{ "} (\gamma _ { r  1 }/2) \sigma _
{ 1 }} {(\omega _ { \ell } - \omega _ { 1 })^ { 2} + \gamma _ { 1
}^ { 2}/4}, $$
$$ \sigma _ { 1 } = (\omega _ { \ell } - \omega _ { 1 })
(\omega _ { \ell } - \omega _ { 2 }) + \gamma _ { 1 } \gamma _ { 2
}/4, $$
$$ \sigma _ { 2 } = (\gamma _ { 1 }/2) (\omega _ { \ell } - \omega _ { 2 })-
(\gamma _ { 2 }/2) (\omega _ { \ell } - \omega _ { 1 }).
$$

 In a quantum well in case of plane
monochromatic exciting wave in time representation, the vector of
the electric field looks like
\begin {equation}
\label {eq39} {\bf E} (z, t) = {\bf e} _ { \ell } e^ {-i \omega _
{ \ell } t} E (z) + c. c.,
\end {equation}
where
\begin {eqnarray}
\label {eq40} &&E (z) = [e ^{i \kappa  z} + (F (z)/F (0)) N]  C _
{ 1 }\nonumber\\ &+& [e^ {-i \kappa  z} + (-1)^ {m _ { c } + m _ {
v }} e ^{-i \kappa  d} (F (z)/F (0)) N]  C _ { 2 }~~~~
\end {eqnarray}
is the sum of plane waves $ \exp (\pm i \kappa  z) $, which are
connected with reflection from quantum-well borders. Besides, (as
it follows from a kind of $F (z) $) the field in a quantum well
contains oscillating terms, describing coordinate dependence of
wave functions of electrons and holes. In time representation,
field vectors to the left $ ({\bf E}^ { \ell } (z, t)) $ and the
right $ ({\bf E}^ { r } (z, t)) $ of the quantum well  are:
\begin {eqnarray}
\label {eq41}
 {\bf E}^ { \ell } (z, t) &=& {\bf e} _ { \ell }  e^ {-i  \omega _ { \ell }  t}
\Big [E _ { 0 } e^ { i  \kappa _ { 1 }  z} + C _ { R }  e^ {-i
\kappa _ { 1 }  z} \Big] + c. c.,
\nonumber\\
 {\bf E} ^{ r } (z, t) &=& {\bf e} _ { \ell }
C_T  e^{-i  ( \omega _ { \ell }  t -  \kappa _ { 1 }  z)}  +  c.
c..
\end {eqnarray}

In a limiting case of a homogeneous environment $ (\nu  =  \nu _ {
1 }, \zeta  =  1) $, the electric field (after substitution (29)
and (30)) becomes $ (\kappa  d \neq\ 0) $:
\begin {eqnarray}
\label {eq42}
 &&{\bf E}^ { \ell} (z, t) = {\bf e} _ { \ell } E _ { 0 }
e^ {-i  \omega _ { \ell }  t} \Bigg [ e ^{ i  \kappa  z} -
 i (-1)^ { m _ { c }  +  m _ { v }
 }\nonumber\\&\times&
\Bigg (\frac {{\tilde \gamma} _ { r  1 }/2} {\Omega _ { 1 } + i
\Gamma _ { 1 }/2} + \frac {{\tilde \gamma} _ { r  2 }/2} {\Omega _
{ 2 } + i \Gamma _ { 2 }/2} \Bigg) e^ {i  \kappa  (d-z)} \Bigg] +
 c.c.,~~~~
\end {eqnarray}
\begin {eqnarray}
\label {eq43} &&{\bf E}^ { r} (z, t) = {\bf e} _ { \ell } E _ { 0
} e^ {-i  \omega _ { \ell }  t}\nonumber\\&\times& \Bigg [1-i
\Bigg (\frac {{\tilde \gamma} _ { r  1 }/2} {\Omega _ { 1 } + i
\Gamma _ { 1 }/2} + \frac {{\tilde \gamma} _ { r  2 }/2} {\Omega _
{ 2 } + i \Gamma _ { 2 }/2} \Bigg) \Bigg] e ^{i  \kappa  z},~~~~~~
\end {eqnarray}
\begin {eqnarray}
\label {eq44} &&{\bf E} (z, t) = {\bf e} _ { \ell } E _ { 0 } e
^{-i  \omega _ { \ell }  t}\nonumber\\&\times&
 \Bigg
[e ^{ i  \kappa  z}
 - i (-1)^ { m _ { c }  +  m _ { v } } e^
{ i  \kappa  d}\nonumber\\&\times& \Bigg (\frac {{\tilde \gamma} _
{ r  1 }/2} {\Omega _ { 1 } + i \Gamma _ { 1 }/2} + \frac {{\tilde
\gamma} _ { r  2 }/2} {\Omega _ { 2 } + i \Gamma _ { 2 }/2} \Bigg)
\frac {F (z)} {F (o)} \Bigg]  +  c.c..~~~~~~
\end {eqnarray}

If, on the contrary, to neglect by the spatial dispersion of the
light wave $ (\kappa  d=0) $, but to consider environment  as not
uniform $ (\zeta \neq 1) $, for allowed transitions $m _ { c } = m
_ { v } $,
\begin {eqnarray}
\label {eq45} {\bf E}^ { \ell} (z, t) &=& {\bf e} _ { \ell } E _ {
0 } e^ {-i  \omega _ { \ell }  t} \Big [e^ { i  \kappa _ { 1 } z}
 +  \frac {\zeta  N} {1 (\zeta-1)
N} \Big]  +  c.c., \nonumber\\
 {\bf E}^ { r} (z, t) &=& {\bf e} _ { \ell } E _ { 0 } e^ {-i  \omega _ { \ell }  t}
\Big [\frac {1+N} {1 (\zeta-1) N}  \Big]  e^ { i  \kappa _
{ 1 }  z}  +  c.c., \nonumber\\
 {\bf E} (t) &=& {\bf e} _ { \ell } E _ { 0 } e^
{-i  \omega _ { \ell }  t} \Big [\frac {1+N} {1 (\zeta-1) N} \Big]
 +  c.c..
\end {eqnarray}

Functions $ \zeta  N / (1- (1 - \zeta)  N) $ and $ (1+N) / (1 -(1
- \zeta)  N) $ in a considered limiting case look like:
$$ \frac {\zeta  N} {1- (1 - \zeta)  N} = - \frac {
i \zeta (\gamma _ {r  1} {\tilde \omega} _2 + \gamma _ {r  2}
{\tilde \omega} _1)/2} {{\tilde \omega} _1 {\tilde \omega} _2+i
\zeta (\gamma _ {r  1} {\tilde \omega} _2 + \gamma _ {r  2}
{\tilde \omega} _1)/2} $$ and
$$ \frac {1+N} {1-(1 - \zeta)  N} = \frac {{\tilde \omega} _1 {\tilde \omega} _2}
{{\tilde \omega} _1 {\tilde \omega} _2+i \zeta (\gamma _ {r  1}
{\tilde \omega} _2 + \gamma _ {r  2} {\tilde \omega} _1)/2}. $$

It is visible that $ \gamma _ { r i} $ and $ \zeta $ enter only in
the form of product $ \gamma _ { r
 i}  \zeta $. It means that at $ \nu \neq \nu_1 $ in case of
narrow quantum wells, the refraction coefficient of barriers $
\nu_1 $ enters in radiative damping, and a refraction coefficient
of quantum well  $ \nu $ does not appear anywhere. Physical sense
of this result is clear: At $ \kappa  d \ll 1 $, it is possible to
pass to the limit $d \to 0 $, when the substance of a quantum well
is absent, but the induced current, corresponding to transitions
with an exciton creation, is kept. Thus, it is proved that
obtained earlier results for narrow quantum wells are true and at
$ \nu \neq \nu_1 $, since formulas include only refraction
coefficients of barriers.

In this limiting case, the field in a quantum well does not depend
on coordinate, since in dipole approximation, the phase of a light
wave does not vary in a quantum well.

\section{Optical characteristics of a quantum well}

In this Section, formulas for the reflection, transmission and
absorption of a plane monochromatic electromagnetic wave by a
quantum well are represented in the general case, when $ \nu \neq
\nu _ { 1 } $ and $ \kappa d \neq 0 $, and in limiting cases of a
homogeneous environment and $ \kappa d = 0 $.

The reflection coefficient $R $ is defined by standard way as the
ratio of the module of the reflected energy flux to the module of
the incident energy flux:
\begin {equation}
\label {eq46} R = {| C _ { R } |^ { 2}\over |E _ { 0 } |^ { 2}}.
\end {equation}

Similarly, the transmission coefficient $T $ is defined as
\begin {equation}
\label {eq47} T = {| C _ { T } |^ { 2} \over |E _ { 0 } |^ { 2}}.
\end {equation}
The dimensionless absorption coefficient $A $ (a share of energy
absorbed by a quantum well), according to (46) and (47), looks
like
\begin {equation}
\label {eq48} A = 1 - R - T.
\end {equation}

Using (31), (32), and (30), it is possible to represent the
reflection $R $ in the form
\begin {equation}
\label {eq49} R = {v _ { 1 } + (L ^{ 2} + G^ { 2})  X _ { 1 }- Y
 L -  Z _ { 1 }  G \over | \Delta |^ { 2}}.
\end {equation}
The denominator $ \Delta $, defined by (21), is been transformed
to
\begin {equation}
\label {eq50} | \Delta |^ { 2} = v + (L ^{ 2} + G ^{ 2})  X - Y L
+ Z  G.
\end {equation}
Functions $L $ and $G $ determine the frequency dependence of
reflection in the region of interband transitions into two excited
energy levels:
\begin {equation}
\label {eq51} L = \sum _ {j=1} ^2 \frac {({\tilde \gamma} _ { r j
}/2)  \Omega _ { j }} {\Omega _ { j }^ { 2}  +  \Gamma _ { j } ^{
2}/4},~~~~~~ G = \sum _ {j=1} ^2 \frac {{\tilde \gamma} _ { r  j }
\Gamma _ { j }/4} {\Omega _ { j }^ { 2}  +  \Gamma _ { j }^ {
2}/4},
\end {equation}
\begin {eqnarray}
\label {eq52}
 L^ { 2} &+& G^ { 2} \equiv  | N |^ 2 =
\frac {({\tilde \gamma} _ { r  1 }/2)^ { 2 }} {\Omega _ { 1 } ^{
2}  +\Gamma _ { 1 } ^{ 2}/4} + \frac {({\tilde \gamma} _ { r  2
}/2)^ { 2 }} {\Omega _ { 2 }^ { 2}  +  \Gamma _ { 2 }^ { 2}/4}
\nonumber\\ &+&
 \frac {2 ({\tilde \gamma} _ { r  1 }/2)
({\tilde \gamma} _ { r  2 }/2) (\Omega _ { 1 } \Omega _ { 2 }  +
 \Gamma _ { 1 }  \Gamma _ { 2 })} {(\Omega _ { 1
}^ { 2}  +  \Gamma _ { 1 }^ { 2}/4) (\Omega _ { 2 }^ { 2}  +
\Gamma _ { 2 }^ { 2}/4)}.
\end {eqnarray}
Coefficients $v, v _ { 1 }, X, X _ { 1 }, Y, Z $ and $Z _ { 1 } $
depend on parameters $ \zeta = \nu / \nu _ { 1 } $ and $ \kappa d
$:
\begin {eqnarray}
\label {eq53} v &=& 4 [4 \zeta ^{ 2} \cos ^{ 2} \kappa d + (\zeta^
{ 2} + 1)^ { 2} \sin ^{ 2} \kappa d],\nonumber\\ v _ { 1 } &=& 4
(\zeta^{ 2} - 1)^ { 2} \sin ^{ 2} \kappa d,
\end {eqnarray}
\begin {eqnarray}
\label {eq54}
 X &=& 8 (\zeta - 1)^ { 2} [\zeta^ { 2} + 1 +
(-1)^ {m _ { c } + m _ { v }} (\zeta^ { 2} - 1) \cos
\kappa d], \nonumber\\
 X _ { 1 } &=& 8 [\zeta^ { 4} + 1 + (-1)^ {m _ {
c } + m _ { v }} (\zeta ^{ 4} - 1) \cos \kappa d],
\end {eqnarray}
\begin {eqnarray}
\label {eq55} Y &=& 8 (\zeta^ { 2} - 1) [(-1)^ {m _ { c } + m _ {
v }} (\zeta^ { 2} +1)\nonumber\\ &+& (\zeta ^{ 2} - 1) \cos \kappa
d ] \sin \kappa d,
\end {eqnarray}
\begin {eqnarray}
 \label {eq56}
Z &=&8 (\zeta - 1) [(-1)^ {m _ { c } + m _ { v }} (\zeta^2+1)+
(\zeta^2-1) \cos \kappa d]\nonumber\\&\times&  [1-(-1)^ {m _ { c }
+ m _ { v }} (\zeta-1)+(\zeta+1) \cos \kappa d]
\nonumber\\
 Z _ { 1 } &=& 8\ (\zeta ^{ 2} - 1)^
{ 2} \sin^ { 2} \kappa d.
\end {eqnarray}

If, in a quantum well, there is one excited energy level, it is
necessary to pass in the formula (49) to the limit $ \gamma _ { r
 2 } \to 0, \gamma _ { r  1 } = \gamma _ { r },
\gamma _ { 1 } \to \gamma, \omega _ { 1 } = \omega _ { 0 } $. In
this case, the reflection coefficient looks like:
\begin {eqnarray}
\label {eq57}
 &&R = \frac {1} {| \Delta |^ { 2}} \Bigg [v _ { 1 } +
\frac {({\tilde \gamma} _ { r }/2)^ { 2} X _ { 1 }- ({\tilde
\gamma} _ { r }/2) (Y \Omega+Z _ { 1 } \Gamma/2)}
{\Omega^ { 2} + \Gamma ^{ 2}/4} \Bigg], \nonumber\\
 &&| \Delta |^ { 2} =v + \frac {({\tilde \gamma} _ { r }/2)^ { 2} X-
({\tilde \gamma} _ { r }/2) (Y \Omega-Z \Gamma/2)} {\Omega^ {
2} + \Gamma ^{ 2}/4}, \nonumber\\
 &&\Omega = \omega _ { \ell } - \omega _ { 0 }
- \varepsilon ^" \gamma _ { r }/2,\nonumber\\&& \Gamma = \gamma +
{\tilde \gamma} _ { r }, {\tilde \gamma} _ { r } = \varepsilon ^{
' } \gamma _ { r }.~~~~
\end {eqnarray}

Since refraction coefficients of barriers and  quantum well, as a
rule, do not differ enough (i.e., $ \zeta \simeq 1 $), in the
denominator $ | \Delta |^ { 2} $, ~~ $v \simeq 4 $ plays a
dominating role. Other multipliers contain factors $ \zeta ^{ 2} -
1 $ or $ (\zeta - 1)^ { 2} $ and are small in comparison with $v $
even in the resonance, when $L\simeq 1 $ and $G \simeq 1 $. In
numerator $R $, the value $v _ { 1} \ll 1 $ and the contribution
of other terms is significant. In particular,
 a sign-variable term $ \sim L $, whose frequency dependence
  is similar to a curve of an anomalous dispersion, plays the essential role.
Such dependence takes place in the refraction coefficient in the
region of frequencies corresponding to absorption peaks.

The transition coefficient $T $ is equal
\begin {equation}
\label {eq58} T = {16 \zeta^ { 2} \ [\ L^ { 2} + (1 - G)^ { 2} ]
\over | \Delta |^ { 2}}
\end {equation}
and does not contain sign-variable terms in the numerator. The
same takes place and for absorption
\begin {eqnarray}
\label {eq59} A = 16 \zeta\ [ \zeta^ { 2} + 1 + (-1)^ {m _ { c } +
m _ { v }} (\zeta^ { 2} - 1) \cos \kappa d] \nonumber\\\times { G
- ( L ^{ 2} + G^ { 2} )  \over | \Delta | ^{ 2}}.
\end {eqnarray}
For one excited energy level, we obtain
\begin {eqnarray}
\label {eq60} T &=& \frac {16 \zeta ^{ 2}} {| \Delta |^ { 2}}
\frac {\Omega^ { 2} + \gamma^ { 2}/4} {\Omega ^{ 2} + \Gamma^
{ 2}/4},\nonumber\\
 A &=& \frac {16 \zeta [\zeta^ { 2} +1 +
(\zeta^ { 2}-1) \cos \kappa d]} {| \Delta |^ { 2}} \frac {\gamma
{\tilde \gamma} _ { r }/4} {\Omega^ { 2} + \Gamma ^{ 2}/4}.
\end {eqnarray}

In approximation of a homogeneous medium $ (\zeta =1) $,
\begin {eqnarray}
\label {eq61} R&=&L^ { 2} +G ^{ 2},\nonumber\\ T&=&L ^{ 2} +
(1-G)^ { 2},\nonumber\\ A&=&2 [G-(L^ { 2} +G ^{ 2})].
\end {eqnarray}
 There is no the term $ \sim L $ in the reflection, i.e., this
approximation does not take into account the contribution into the
reflection, originating from sign-variable terms in the current
(8).

In a limiting case $ \kappa d =0 $,
\begin {equation}
\label {eq62}
| \Delta |^ { 2} \to | \Delta_0 |^ { 2} =16 \zeta^2+X_0
(L_0^2+G_0^2) +Z_0  G_0,
\end {equation}
%
%
\begin {eqnarray}
\label {eq63} R&=&{X _ {10} (L_0^2+G_0^2) \over |\Delta_0 |^ {
2}},\nonumber\\
 T&=&{16 \zeta ^ 2 [L _ { 0 }^ { 2} +
(1-G _ { 0 }) ^{ 2}] \over |\Delta _ { 0 } |^ {
2}},\nonumber\\
 A&=&16 \zeta [\zeta^2+1 + (-1)^ {m _ { c } + m _
{ v }} (\zeta^2-1)]  \nonumber\\
&\times&{ G _ { 0 }-(L _ { 0 }^ { 2} +G _ { 0 }^ { 2})\over |
\Delta _ { 0 } |^ { 2}}.
\end {eqnarray}

 Coefficients $X_0 $, $Z_0 $ and $X _ {10} $ originate from (54)
and (56) if $ \kappa d = 0 $:
$$ X_0=8 (\zeta-1) ^2 [\zeta^2+1 + (-1)^ {m _ { c } + m _ { v }}
(\zeta^2-1)],$$
$$ X _ {01} =8 [\zeta^4+1 + (-1)^ {m _ { c } + m
_ { v }} (\zeta^4-1)], $$
$$ Z_0=8 (\zeta-1)  [\zeta^2-1 + (-1)^ {m _ { c } + m _ { v }} (\zeta^2+1)]
$$
$$\times [\zeta+1-(-1)^ {m _ { c } + m _ { v }}
(\zeta-1)].
$$

Functions $G _ { 0 } $ and $L _ { 0 } $ differ from $G $ and $L $
by replacement of $ {\tilde \gamma} _ { r  j } $ by $ \gamma _ { r
 j } $, since at $ \kappa d \to 0, ~~ \varepsilon { ' }
\to 1 $. Besides,  $ \varepsilon { " } = 0 $ in $ \Omega _ { j } $
(37), and  $ \varepsilon =1 $ in $ \beta _ { j } $ (34).

 \section{ Light reflection, absorption and transmission  for arbitrary $ \Phi (z) $}

In Sections 2 and 3, coefficients $R $, $T $ and $A $ were
obtained for $ \Phi (z) $ in the form of (4), that is true for
free electron-hole pairs. In this Section, $ \Phi (z) $ is
considered as arbitrary real function, in particular, it may be an
excitonic function. Calculation, similar to Section 2, leads to
following expressions for constants $C_R $ and $C_T $:
\begin {eqnarray}
\label {eq64} C_R &=& {E_0 \over \Delta_1} \big\{2i (\zeta^2-1)
\sin
\kappa d \nonumber\\
&+&  2 \big [e^ {-i \kappa  d}  w_R  +  (-1) ^{m _ { c } + m _ { v
}} (\zeta^2-1) \big] N \big \},
\end {eqnarray}
\begin {equation}
\label {eq65} C_T = {4E_0 \over \Delta_1} e ^{-i \kappa_1  d}
\zeta
 \big [1  +  (-1)^ {m _ { c } + m _ { v }} e^ {-i
\kappa  d}  N \big],
\end {equation}
where
\begin {eqnarray}
\label {eq66} \Delta_1 &= &(\zeta+1) ^2e ^{-i \kappa  d}  -
(\zeta-1) ^2e^ {i \kappa  d}-  2 (\zeta-1) \nonumber\\&\times&
\big [(\zeta+1) e^ {-i \kappa  d}  w  +  (-1)^ {m _ { c } + m _ {
v }} (\zeta-1) \big] N.
\end {eqnarray}
 In (64) - (66), function $N $ is defined by the formula (34) (or
(36)), and constants $w $ (real) and $w_R $ (complex) are:
\begin {equation}
\label {eq67} w = {(-1)^ {m _ { c } + m _ { v }}\over 2|F (0) |
^2} \Big[F ^2 (0) e^ {-i \kappa  d}  +  [F^* (0) ] ^2  e ^{i
\kappa d}\Big] ,
\end {equation}
\begin {eqnarray}
\label {eq68} &&w_R={(-1)^ {m _ { c } + m _ { v }}\over 2|F (0) |
^2 d}\nonumber\\
&\times& \Big[(\zeta+1) ^2F^2(0)  e^{-i \kappa d}+ (\zeta-1) ^2
[F^* (0) ] ^2 e ^{i \kappa d}\Big].~~~~
\end {eqnarray}
At obtaining of these formulas, relationships
 $F (d) =  e^{i \kappa  d} F^ * (0),   Re \varepsilon  =
\varepsilon  =  |F (0) | ^2 $, which follow from (19), were used
($ \Phi (z) $ is considered real since represents one-dimensional
movement). If (4) is used for $ \Phi (z) $, then $w \to 1, w_R \to
\zeta^2+1 $ and formulas (64) - (66)) coincide with (30) and (31).

It is visible from comparison of expressions (64) - (66) with
coefficients $C_R, C_T $ and $ \Delta $, obtained in Section 2,
that the structure of the function $N $, which basically
determines the frequency dependence, does not depend on type of $
\Phi (z) $. Only values $Re ~ \varepsilon = \varepsilon ' $  and
$Im~ \varepsilon = \varepsilon " $ depend on type of $ \Phi (z) $,
i.e., the factor at $ \gamma_r $ and the shift of resonant
frequencies. Factors $w $ and $w_R $, entering in $ \Delta_1 $ and
$C_R $, can change only a ratio between contributions in optical
characteristics of values $L $ and $G $. Therefore, changes of
optical characteristics will be hardly radical due to changes of $
\Phi (z) $.

 \section{Discussion of results}

Frequency dependencies of the light reflection $R $ and
transmission $T $ are calculated for a case $m_c=m_v=1 $ and $
\gamma _ {r1} = \gamma _ {r2} = \gamma_r, \gamma_1 = \gamma_2 =
\gamma $, that corresponds to the system of the polaron $A $ and a
hole with the Landau quantum number $n=1 $ in conditions of the
magnetophonon resonance \cite{23}. Calculations were spent with
the help of formulas (46) and (47). Functions $C_R $ and $C_T $
from (30) and the denominator $ \Delta $ from (31) are used, as
well as function $N $ (34).

In Fig.1 and Fig.2, the reflection $R $ is shown for small values
of the ratio $ \gamma_r / \gamma $. It is visible that curves 3
and 4 for the case $ \zeta = \nu / \nu_1 \neq 1 $ and $ \kappa d
\neq 0 $ differ considerably from the case of $ \zeta = 1 $ (a
curve 2) at the same value $ \kappa d $. As it was mentioned
above, the formula for the current (8) contains sign-variable
terms $ \sim (\omega_ \ell - \omega _i) $ and terms $ \sim
\gamma_i [(\omega_ \ell - \omega _i) ^2 + \gamma_i^2] $,
corresponding to  absorption curve. Curves 3 and 4 concern to a
case, when
 sign-variable terms dominate. Therefore, $R
(\omega_ \ell) $ is similar to the curve of anomalous dispersion
in this case. Alongside with changes of the form of the reflection
curve,  the increase of $R $ (in 15 times approximately in Fig.1
and twice in Fig.2 ) rakes place in comparison with cases $ \kappa
d \neq 0, \zeta = 1 $ and $ \kappa d = 0, \zeta \neq 1 $.
 \begin{figure}
 \includegraphics [] {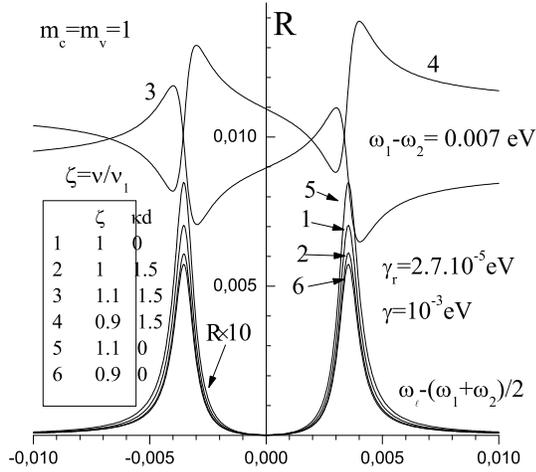} 
 \caption[*]{\label{Fig.1.eps}Dependence of light reflection $R $ on
an exciting wave frequency $ \omega_ \ell $ for a case $ \gamma_r
/ \gamma = 0.027 $. Curves 1,2,5,6 are 10 times increased. The
table of designations concerns also to Figs.2 - 6.}
 \end{figure}
 \begin{figure}
 \includegraphics [] {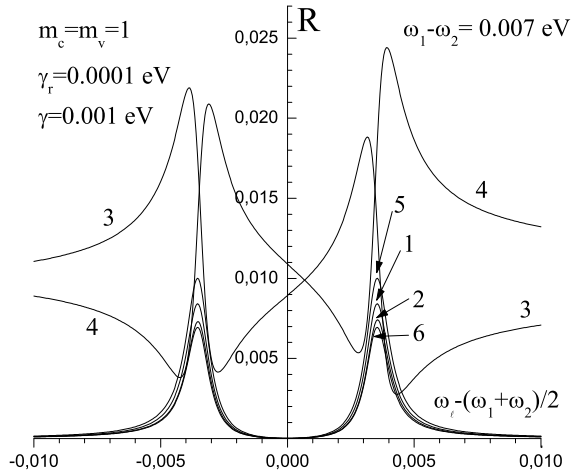} 
 \caption[*]{\label{Fig.2.eps}Same as Fig.1 for $ \gamma_r / \gamma = 0.1 $.}
 \end{figure}

 At $\gamma_r / \gamma = 1 $, influence of the absorption  dominates, as
it is visible in Fig.3. However, influence of sign-variable terms
is still noticeable: Curves 3 and 4 ($ \zeta = 1.1 $ and $ \zeta =
0.9 $, respectively) differ from curve 2 ($ \zeta = 1 $). At the
further increase of $ \gamma_r / \gamma $, the influence of the
anomalous dispersion becomes practically imperceptible and
functions $R, T $ and $A $ coincide with obtained in \cite{22}.
Curves of the transmission $T $ are shown in Figs.4-6 for the same
magnitudes of parameters $ \kappa d,~~ \zeta $ and $ \gamma_r /
\gamma $, as curves in Figs.1-3. It is visible that transmission
is poorly sensitive to changes of $ \zeta $. It happens due to
absence in $T (\omega_ \ell) $ terms proportional to $ \omega_
\ell - \omega_1 $ and $ \omega_ \ell - \omega_2 $ (see (58)). A
small distinction is determined by the denominator $ | \Delta |^2
$ which contains terms linear on $ \omega_ \ell - \omega_j $. The
absorption $A $, in general, is poorly sensitive to changes of
parameter $ \zeta $ in all area of changes of $ \gamma_r / \gamma
$.
 \begin{figure}
 \includegraphics [] {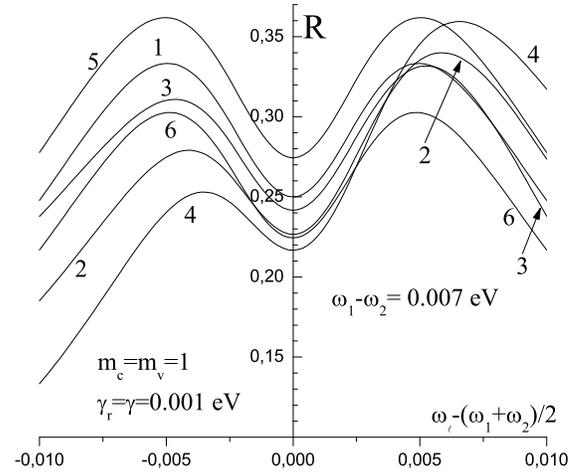} 
 \caption[*]{\label{Fig.3.eps}Same as Fig.1 for $ \gamma_r = \gamma $.}
 \end{figure}
 \begin{figure}
 \includegraphics [] {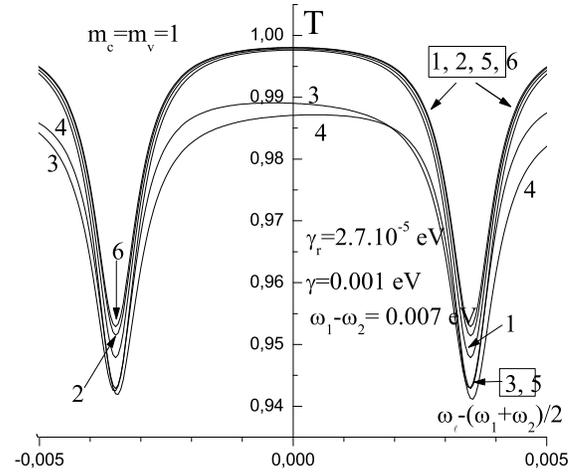} 
 \caption[*]{\label{Fig.4.eps}Dependence of transmission $T $ on frequency $ \omega_ \ell $ of an exciting
wave for a case $ \gamma_r / \gamma = 0.027 $.}
 \end{figure}

 $R $ and $T $ for a case $ \zeta = 1,
\kappa d = 1.5 $ (curve 2) are shown also in figures. Some
asymmetry of the curve 2 in comparison with the case $ \kappa d =
0, \zeta \neq 1 $ (curves 1, 5 and 6) attracts our attention. This
asymmetry is a consequence of the account in the theory of supreme
orders on interaction of an electromagnetic wave with the
electronic system that leads to occurrence of shifts of resonant
frequencies. These shifts combine with resonant frequencies $
\omega_1 $ and $ \omega_2 $. At transition to the limit $ \kappa d
\to 0 $, shifts disappear and symmetry of curves is restored.

Finally, curves 5 and 6 in figures correspond to the case $ \kappa
d = 0, \zeta \neq 1 $. Curves are symmetric here, and influence of
parameter $ \zeta $ is shown in the form of a parallel shift of
curves (if $ \gamma_r = \gamma $), or changes in the vicinity of
extrema ($ \gamma_r \leq \gamma $).
 \begin{figure}
 \includegraphics [] {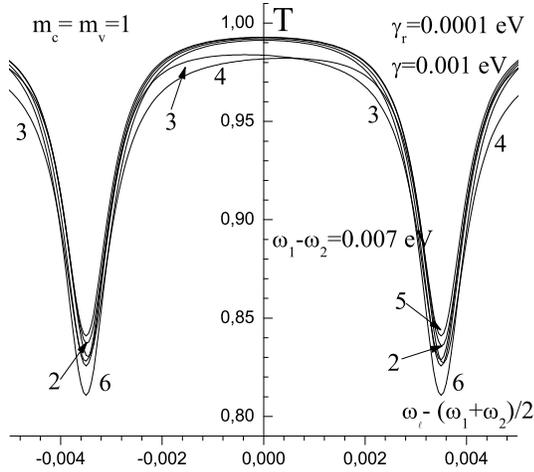} 
 \caption[*]{\label{Fig.5.eps}Same as Fig.3 for $ \gamma_r / \gamma = 0.1 $.}
 \end{figure}
\begin{figure}
\includegraphics[]{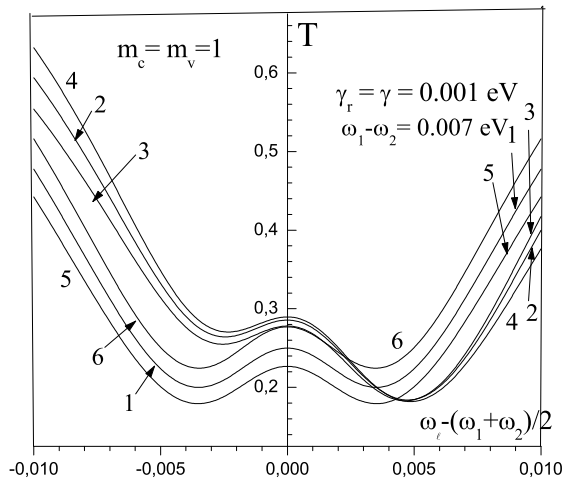} 
\caption[*]{\label{Fig.6.eps}Same as Fig.3 for $ \gamma_r= \gamma
$.}
\end{figure}
The main conclusion of our calculations consists that at greater
radiative lifetimes of excited states in comparison with
nonradiative lifetimes (which are determined, in particular, by
electron and hole scattering on impurities and phonons) frequency
dependence of reflection is determined basically by sign-variable
terms in the expression for the current density. In this case, it
is possible to neglect by $ \gamma_r $ in resonant denominators,
i.e., to solve the problem in linear approximation on interaction
of an electromagnetic wave with the electronic system.

\end {document}